\documentclass{moriond}

\def\be{\begin{equation}}
\def\ee{\end{equation}}
\def\bea{\begin{eqnarray}}
\def\eea{\end{eqnarray}}

\newcommand{\Photo}{}

\begin{document}

%------------------------BEGINNING OF NIKHEF TITLE PAGE----------------------

\thispagestyle{empty}

\begin{flushright}
Nikhef-2018-019\\
\end{flushright}

\vspace{2.0truecm}
\begin{center}
\boldmath
\large\bf New Probes of New Physics with Leptonic Rare B Decays
\unboldmath
\end{center}

\vspace{0.9truecm}
\begin{center}
 Robert Fleischer\,${}^{a,b}$\\[0.1cm]

${}^a${\sl Nikhef, Science Park 105, 1098 XG Amsterdam, Netherlands}

${}^b${\sl  Department of Physics and Astronomy, Faculty of Science, Vrije Universiteit Amsterdam,\\
1081 HV Amsterdam, Netherlands}
\end{center}

\vspace{2.9truecm}

\begin{center}
{\bf Abstract}
\end{center}

{\small
\vspace{0.2cm}\noindent
Decays of the kind $B^0_{s,d}\to\ell^+\ell^-$ belong to the most favourable processes for probing the flavour structure of the Standard Model, with outstanding sensitivity to new (pseudo)-scalar contributions. While the branching ratio of $B^0_s\to\mu^+\mu^-$ has already been measured at the LHC in the ballpark of the Standard Model expectation, there is still significant room for New-Physics effects. We discuss how these may be revealed in the future super-high precision era of $B$-decay studies by utilising new theoretically clean observables, including CP-violating asymmetries. Another promising decay is $B^0_{s}\to e^+e^-$, which has received little attention in view of its enormously helicity suppressed Standard Model branching ratio, with the most recent experimental upper bound dating back to 2009. Using the current constraints on New Physics from $B^0_s\to\mu^+\mu^-$ as a guideline, we show that the $B^0_s\to e^+e^-$ branching ratio may be hugely enhanced through new (pseudo)-scalar contributions up to the regime of $B^0_{s}\to\mu^+\mu^-$.}

\vspace{3.9truecm}

\begin{center}
{\sl Invited talk at Rencontres de Moriond 2018, QCD and High Energy Interactions\\
La Thuile, Italy, 17--24 March 2018\\
To appear in the Proceedings}
\end{center}

\vfill
\noindent
May  2018

\newpage
\thispagestyle{empty}
\vbox{}
\newpage
 
\setcounter{page}{1}

%------------------------END OF NIKHEF TITLE PAGE------------------------------

\vspace*{4cm}
\title{NEW PROBES OF NEW PHYSICS WITH LEPTONIC RARE B DECAYS}

\author{R. FLEISCHER}

\address{Nikhef, Science Park 105, 1098 XG Amsterdam and  
Department of Physics and Astronomy,\\ Faculty of Science, Vrije Universiteit Amsterdam, 1081 HV Amsterdam, Netherlands}

\maketitle\abstracts{Decays of the kind $B^0_{s,d}\to\ell^+\ell^-$ belong to the most favourable processes for probing the flavour structure of the Standard Model, with outstanding sensitivity to new (pseudo)-scalar contributions. While the branching ratio of $B^0_s\to\mu^+\mu^-$ has already been measured at the LHC in the ballpark of the Standard Model expectation, there is still significant room for New-Physics effects. We discuss how these may be revealed in the future super-high precision era of $B$-decay studies by utilising new theoretically clean observables, including CP-violating asymmetries. Another promising decay is $B^0_{s}\to e^+e^-$, which has received little attention in view of its enormously helicity suppressed Standard Model branching ratio, with the most recent experimental upper bound dating back to 2009. Using the current constraints on New Physics from $B^0_s\to\mu^+\mu^-$ as a guideline, we show that the $B^0_s\to e^+e^-$ branching ratio may be hugely enhanced through new (pseudo)-scalar contributions up to the regime of $B^0_{s}\to\mu^+\mu^-$.}

\section{Setting the Stage}
\begin{figure}[htbp] %  figure placement: here, top, bottom, or page
 \centering
\includegraphics[width=6.0truecm]{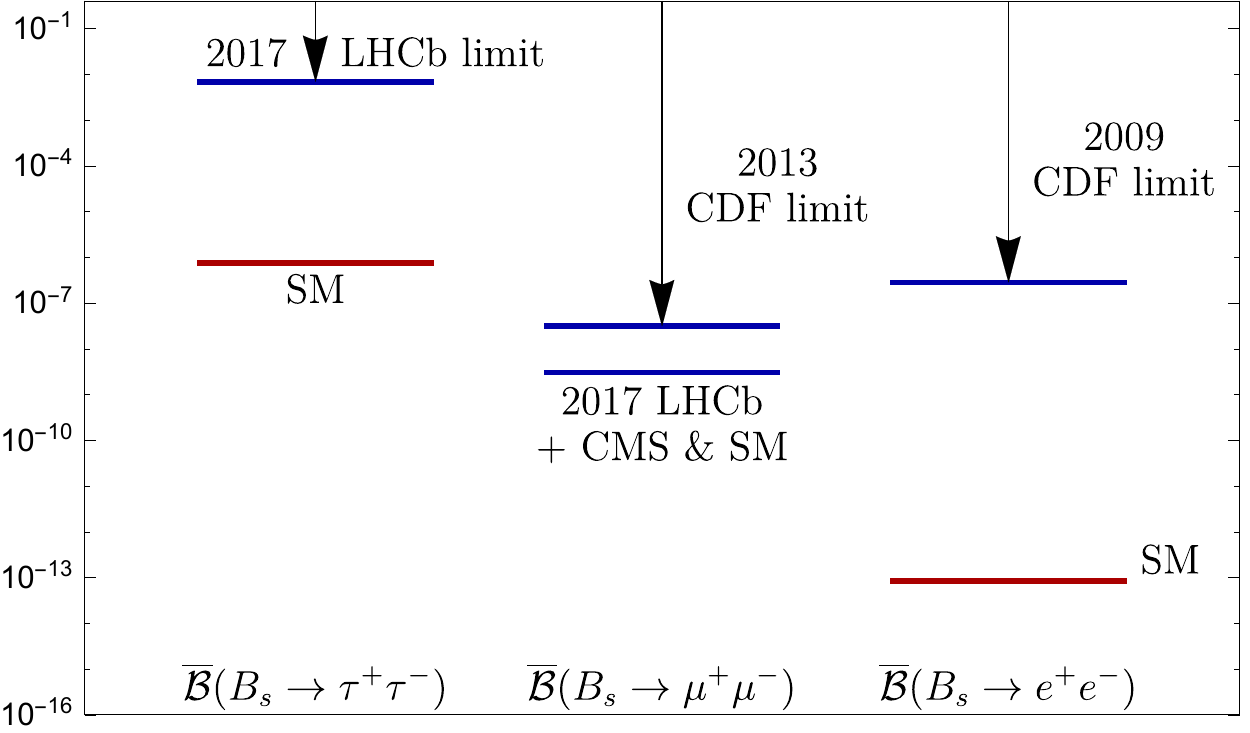} \quad
\includegraphics[width=6.0truecm]{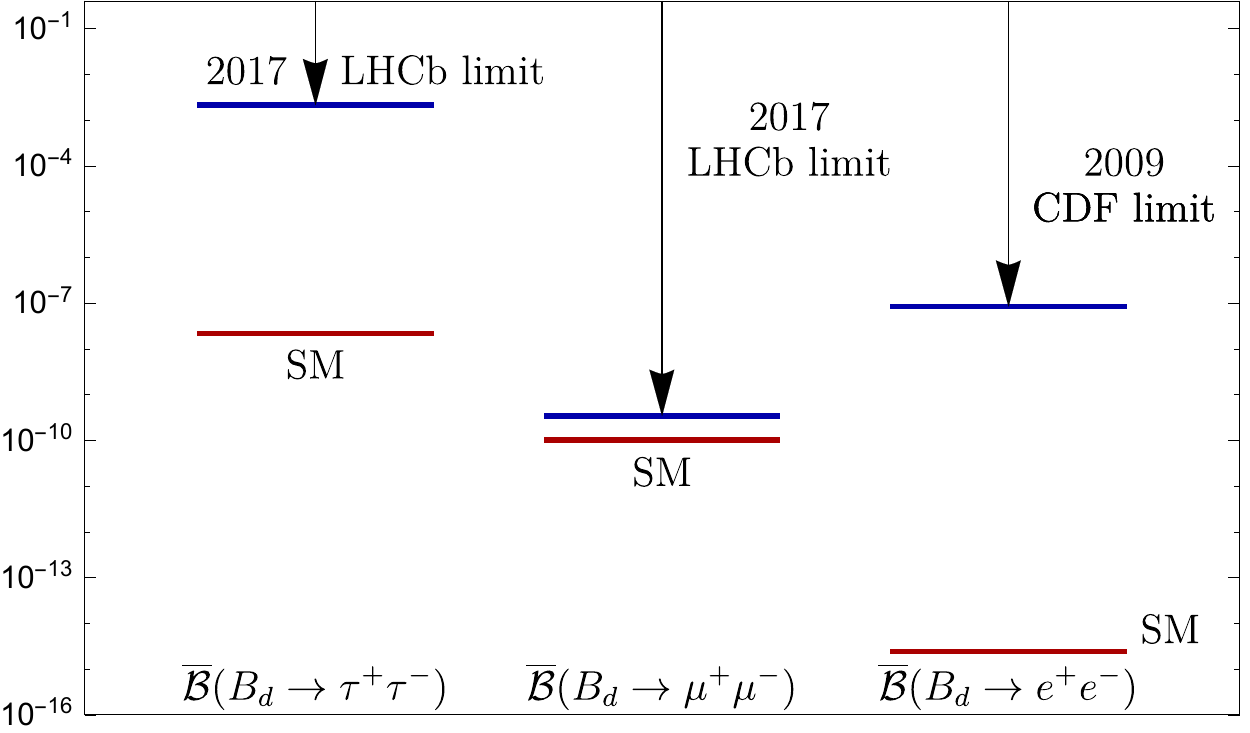} 
\caption{Overview of experimental information on $B_q\to\ell^+\ell^-$ branching ratios as defined in Eq.\ (\ref{BR-def}) and 
comparison with the corresponding SM predictions.}\label{fig:1}
 \end{figure}

Within the Standard Model (SM), the leptonic decays $B^0_q\to\ell^+\ell^-$ ($q=s,d$) receive only loop contributions from penguin 
and box topologies, and show a helicity suppression which results in branching ratios proportional to $m_\ell^2$, where $m_\ell$ 
denotes the masses of the final state leptons. Another key feature is the simple situation concerning strong interactions, which are 
described by a single hadronic parameter, the $B_q$ decay constant $f_{B_q}$. These modes belong to the cleanest rare $B$ 
decays and offer an outstanding setting to explore the flavour sector of the SM, with high sensitivity to New Physics (NP) contributions. 
Particularly interesting are new (pseudo)-scalars, which may lift the helicity suppression. In Fig.~\ref{fig:1}, we show a compilation of  
experimental information in comparison with the SM picture. So far, only $B^0_s\to\mu^+\mu^-$ has been observed, which was a 
highlight of LHC run 1. In the case of  $B^0_{s,d}\to\tau^+\tau^-$, the helicity suppression is not very effective due to the 
large $\tau$ mass but the $\tau$ reconstruction makes experimental analyses challenging. Interestingly, the $B_{s,d}\to e^+e^-$ 
modes, which are extremely helicity suppressed in the SM, have not yet received attention at the LHC.

New observables of the decay $B_s\to\mu^+\mu^-$ were pointed out, which offer interesting probes at the high-precision 
frontier.~\cite{PRL} In the following, we focus on the constraints for NP effects following from the current $B_s\to\mu^+\mu^-$ 
data,\cite{NP} their implications for the branching ratios of $B^0_{s,d}\to \tau^+\tau^-$, $B^0_{s,d}\to e^+e^-$,\cite{Bee} and 
address the impact of new sources of CP violation.\cite{CPV-rare}

\section{In Pursuit of New Physics}
The theoretical framework is given by effective quantum field theory, where the decays at hand are described by a low-energy
effective Hamiltonian.\cite{PRL} In the SM, only the operator $O_{10}=(\bar q \gamma_\mu P_L b) (\bar\ell\gamma^\mu \gamma_5\ell)$ 
contributes with a real Wilson coefficient. In the presence of NP, new four-fermion operators involving (pseudo)-scalar lepton
densities may enter. Their effect is described by short-distance coefficients $P^q_{\ell\ell}$ and $S^q_{\ell\ell}$, where the former
includes the SM and pseudo-scalar NP effects while the latter originates from new scalars. In the SM, we have
$P^q_{\ell\ell}=1$ and $S^q_{\ell\ell}=0$.

Due to the presence of $B^0_s$--$\bar B^0_s$ mixing and the sizeable $B_s$ decay width difference 
$\Delta\Gamma_s/\Gamma_s\sim0.1$, a subtle difference arises between the untagged, time-integrated branching ratio 
\begin{equation}\label{BR-def}
\overline{\mathcal{B}}(B_{s}\to\mu^+\mu^-)
	\equiv \frac{1}{2}\int_0^\infty \langle \Gamma(B_s(t)\to \mu^+\mu^-)\rangle\, dt
      \,	\stackrel{{\rm LHC}}{=}\,
 \left(3.0\pm 0.5\right) \times 10^{-9},
\end{equation}
measured at the LHC, and theoretical predictions $\mathcal{B}(B_s \to \mu^+\mu^-)$ which usually refer to a setting without the
oscillations.\cite{PRL,BR-paper} The conversion involves an observable ${\cal A}^{\mu\mu}_{\Delta\Gamma_s}$, which depends 
on $P_{\mu\mu}$ and $S_{\mu\mu}$ and takes the SM value +1, yielding\cite{Bee} 
$\overline{\mathcal{B}}(B_{s}\to\mu^+\mu^-)_{\rm SM}= (3.57\pm0.16)\times 10^{-9}$. Electromagnetic corrections were recently 
calculated in Ref.~6 %\cite{BSS}
and were found to be tiny. 
The observable ${\cal A}^{\mu\mu}_{\Delta\Gamma_s}$ contains information equivalent to the effective lifetime
\begin{equation}
\tau_{\mu\mu} \equiv \frac{\int_0^\infty t\,\langle \Gamma(B_s(t)\to \mu^+\mu^-)\rangle\, dt}
	{\int_0^\infty \langle \Gamma(B_s(t)\to \mu^+\mu^-)\rangle\, dt} =  \left[2.04 \pm 0.44 ({\rm stat}) 
	\pm 0.05 ({\rm syst}) \right]\hbox{ps}
\end{equation}
which was measured by the LHCb collaboration for the first time with the value given above.\cite{LHCb-ADG}

\begin{figure}[bp] %  figure placement: here, top, bottom, or page
 \centering
\includegraphics[width=6.5truecm]{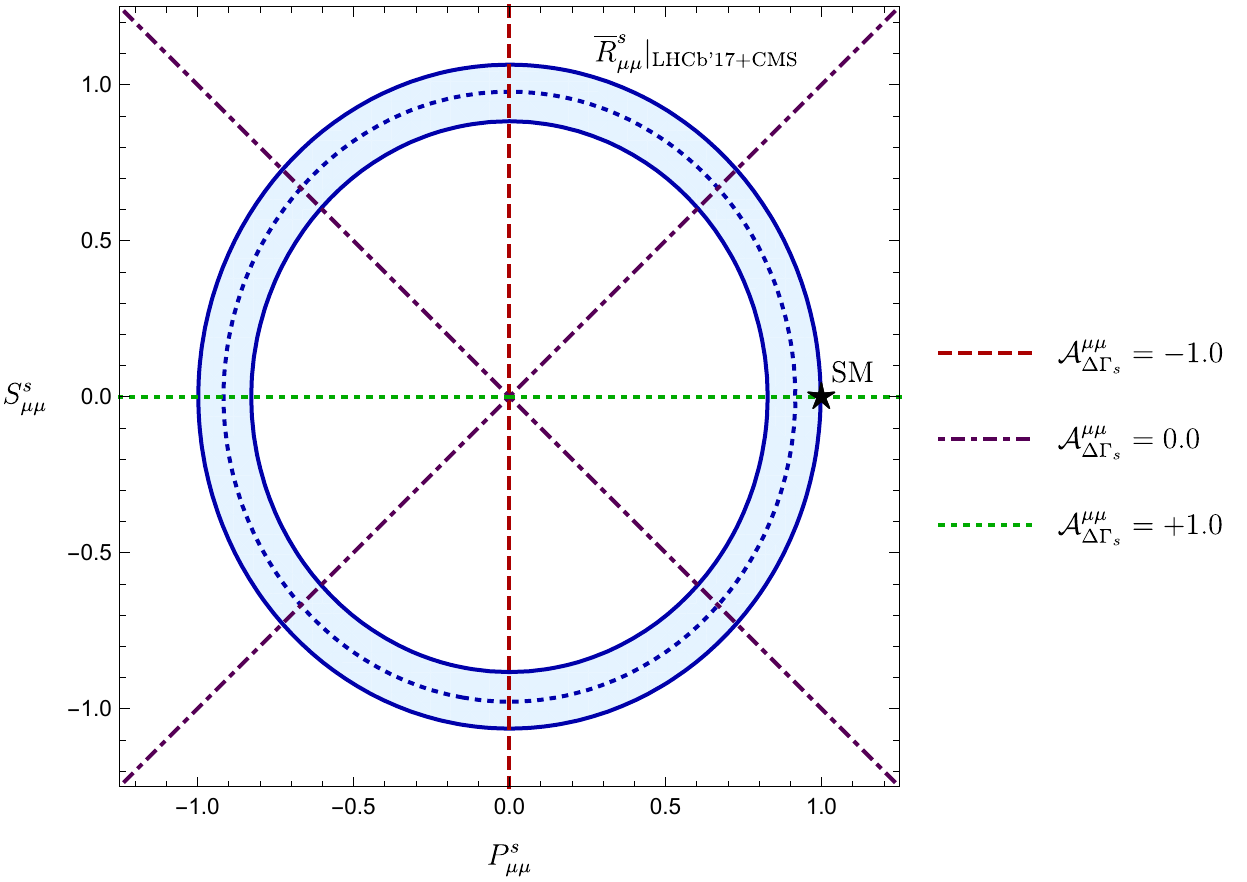} 
\caption{Constraints in the $P_{\mu\mu}$--$S_{\mu\mu}$ plane following from the LHC data and 
impact of ${\cal A}^{\mu\mu}_{\Delta\Gamma_s}$ (see Ref.~3).}\label{fig:2}
 \end{figure}

In order to probe NP effects through the measured $B^0_s\to\mu^+\mu^-$ branching ratio, the quantity
\begin{equation}
 \overline{R}_{\mu\mu}^s \equiv \overline{\mathcal{B}}(B_s\to\mu^+\mu^-)/\overline{\mathcal{B}}
   (B_s\to\mu^+\mu^-)_{\rm SM} = 0.84 \pm 0.16
\end{equation}
plays a central role.\cite{PRL,NP} Assuming real coefficients $P_{\mu\mu}$ and $S_{\mu\mu}$, we obtain the constraints shown in 
Fig.~\ref{fig:2}. Interestingly, $\overline{R}_{\mu\mu}^s$ alone does not allow a separation of these contributions and sizeable NP 
effects could still be present.\cite{Bee} They could be revealed through a future measurement of 
${\cal A}^{\mu\mu}_{\Delta\Gamma_s}$. Unfortunately, the current value of ${\cal A}^{\mu\mu}_{\Delta\Gamma_s}=8.24\pm10.72$ 
does not yet have an impact.

\begin{figure}[htbp] %  figure placement: here, top, bottom, or page
 \centering
\includegraphics[width=12.0truecm]{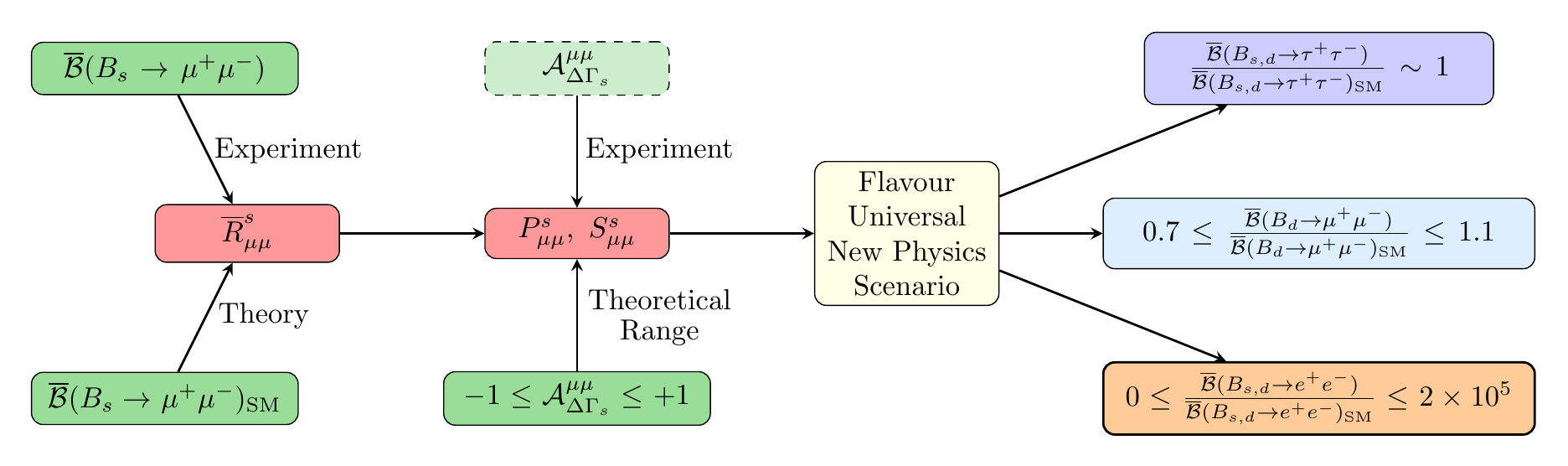} 
\caption{Flowchart to explore the impact of $B_s\to\mu^+\mu^-$ NP constraints for other $B_{s,d}\to\ell^+\ell^-$ decays.}\label{fig:3}
 \end{figure}

\begin{figure}[b] %  figure placement: here, top, bottom, or page
 \centering
\includegraphics[width=6.5truecm]{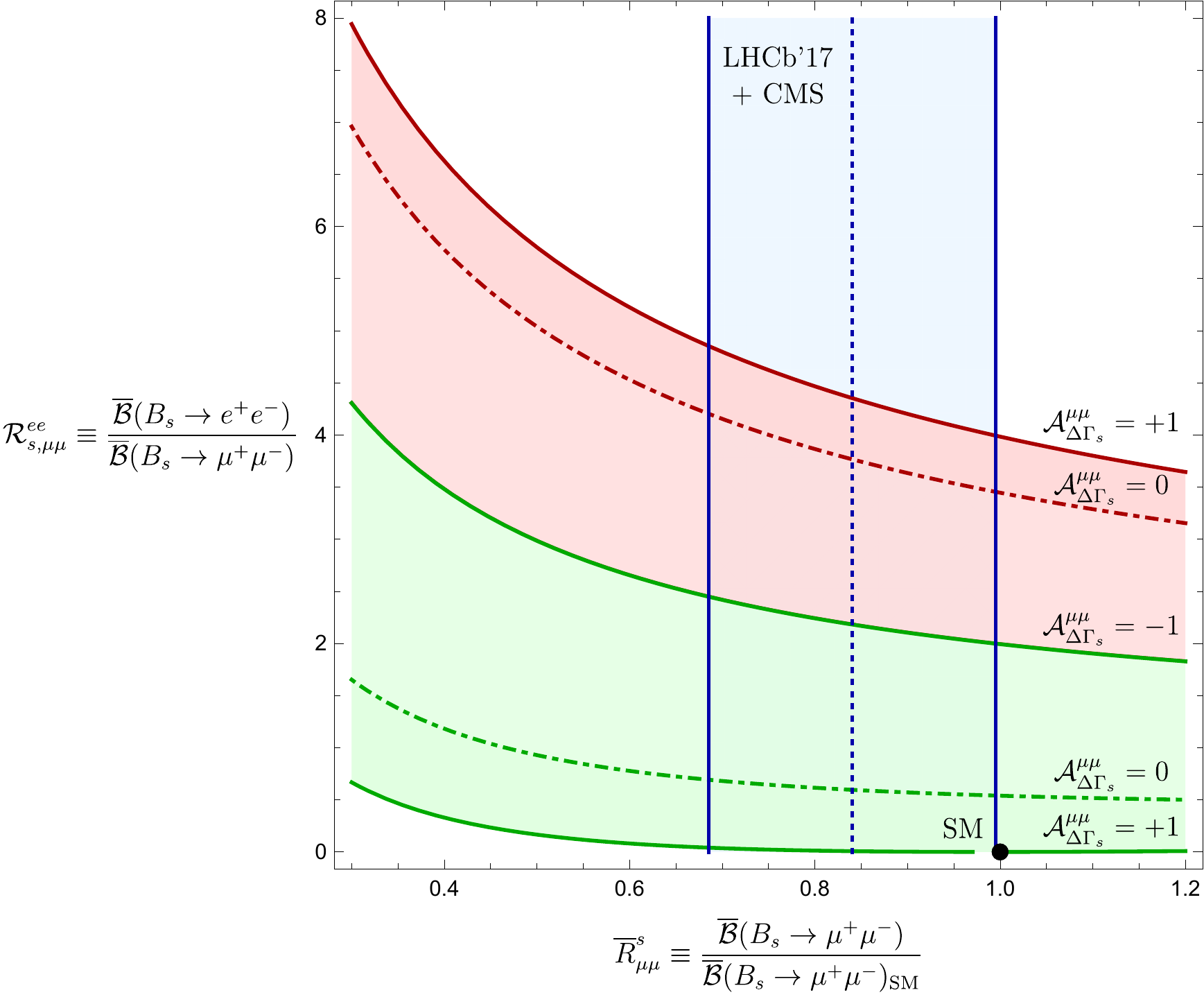} 
\caption{Correlation between the $B_s\to e^+e^-$ and $B_s\to\mu^+\mu^-$ branching ratios in the FUNP scenario.}\label{fig:4}
 \end{figure}

Let us now explore implications of these NP constraints for other $B_q\to\ell^+\ell^-$ processes.\cite{Bee} 
To this end, we employ a scenario with flavour-universal NP (FUNP) contributions, which is characterised by the feature that 
$C_{10}^{\ell\ell(')}$, $C_P^{\ell\ell(')}$, $C_S^{\ell\ell(')}$ do not depend on flavour labels. In Fig.~\ref{fig:3}, the corresponding strategy 
is illustrated in a flowchart. 

In the case of $B^0_d\to\mu^+\mu^-$, the ratio
\begin{equation}
\frac{\overline{\mathcal{B}}(B_d \to \mu^+\mu^-)}{\overline{\mathcal{B}}(B_s \to \mu^+\mu^-)}\propto
\left[\frac{|P^d_{\mu\mu}|^2 + |S^d_{\mu\mu}|^2}{|P^s_{\mu\mu}|^2 + |S^s_{\mu\mu}|^2}\right] 
\left(\frac{f_{B_d}}{f_{B_s}}\right)^2\left|\frac{V_{td}}{V_{ts}}\right|^2 
\end{equation}
is a particularly interesting quantity, where the ratio of CKM matrix elements can be determined from an analysis of the unitarity triangle.
In the FUNP scenario, an essentially linear correlation between the branching ratios arises, with a moderate suppression of 
$\overline{\mathcal{B}}(B_d\to\mu^+\mu^-)$ with respect to the SM expectation, in analogy to the current LHC data for 
$B_s\to\mu^+\mu^-$. 

Concerning $B^0_q\to\tau^+\tau^-$ decays, the NP effects are strongly suppressed by the mass ratio $m_{\mu}/m_{\tau} \sim 0.06$ 
in the FUNP scenario, resulting in 
\begin{equation}
0.8 \leq  \overline{R}_{\tau\tau}^s \equiv \overline{\mathcal{B}}(B_s\to\tau^+\tau^-)/\overline{\mathcal{B}}
   (B_s\to\tau^+\tau^-)_{\rm SM} \leq 1.0,  \quad 0.995 \leq \mathcal{A}_{\Delta\Gamma_s}^{\tau\tau}\leq 1.000,
\end{equation}
with a similar picture for $B^0_d\to\tau^+\tau^-$. First experimental bounds were obtained by LHCb.\cite{LHCb-Btautau}

In the case of $B^0_q\to e^+e^-$, we have a situation complementary to $B^0_q\to\tau^+\tau^-$ within the FUNP 
framework, where the NP effects are hugely amplified by the mass ratio $m_{\mu}/m_e \sim 207$. In this scenario, 
the (pseudo)-scalar New Physics contributions lift the helicity suppression of the extremely small SM branching ratio,
as illustrated in Fig.~\ref{fig:4}, where the red and green bands describe $P^s_{\mu\mu}<0$ and $P^s_{\mu\mu}>0$,
respectively. These results correspond to 
\begin{equation}
0\leq \overline{R}^s_{ee}\leq 1.7\times 10^{5}, \quad
0\leq \overline{\mathcal{B}}(B_s \rightarrow e^+e^-)\leq1.4\times 10^{-8};
\end{equation}
a similar picture arises for the $B_d\to e^+e^-$ decay, with $0\leq\overline{\mathcal{B}}(B_d \rightarrow e^+e^-)\leq 4.0 \times 10^{-10}$.
The most recent experimental constraints on these modes were obtained by the CDF collaboration:
$\overline{\mathcal{B}}(B_s \to e^+ e^-) < 2.8 \times 10^{-7}$ and $\overline{\mathcal{B}}(B_d \to e^+ e^-) < 8.3 \times 10^{-8}$,
and date back to 2009.\cite{CDF-Bsee} It would be most interesting to search for these modes at the LHC, where an observation 
would give an unambiguous signal for physics beyond the SM.

\section{Impact of CP Violation}
New sources of CP violation may enter through phases of the short-distance coefficients. In the case of $B_s\to\mu^+\mu^-$
decays, we have the following time-dependent CP asymmetry:\cite{PRL,NP}
\begin{equation}\label{CP-asym}
\frac{\Gamma(B^0_s(t)\to \mu_\lambda^+\mu^-_\lambda)-
\Gamma(\bar B^0_s(t)\to \mu_\lambda^+\mu^-_\lambda)}{\Gamma(B^0_s(t)\to \mu_\lambda^+\mu^-_\lambda)+
\Gamma(\bar B^0_s(t)\to \mu_\lambda^+\mu^-_\lambda)}
=\frac{{\cal C}_{\mu\mu}^\lambda\cos(\Delta M_st)+{\cal S}_{\mu\mu}\sin(\Delta M_st)}{\cosh(y_st/\tau_{B_s}) + 
{\cal A}_{\Delta\Gamma_s}^{\mu\mu} \sinh(y_st/\tau_{B_s})},
\end{equation}
where $\lambda$ is the muon helicity and $y_s\equiv \Delta\Gamma_s \tau_{B_s}/2$. The 
${\cal C}_{\mu\mu}^\lambda$ term cancels in the helicity-averaged rates, and 
$({\cal C}_{\mu\mu}^\lambda)^2+({\cal S}_{\mu\mu})^2+({\cal A}_{\Delta\Gamma_s}^{\mu\mu})^2=1$. These 
CP asymmetries were analysed within specific NP models,\cite{NP} and a detailed study to probe possible CP-violating phases of 
$P^s_{\mu\mu}\equiv |P^s_{\mu\mu}|e^{i\varphi_{P_s}^{\mu\mu}}$ and  
$S^s_{\mu\mu}\equiv |S^s_{\mu\mu}|e^{i\varphi_{S_s}^{\mu\mu}}$ has recently been performed,\cite{CPV-rare} showing that
the CP asymmetries do not offer sufficient information to determine all parameters from the data. However, assuming specific 
scenarios, much sharper pictures can be obtained. Explorations of CP violation offer valuable insights and are an essential part for 
revealing the full dynamics of the $B^0_s\to\mu^+\mu^-$ decays.

\section{Conclusions}
We are moving towards new frontiers with $B_q\to\ell^+\ell^-$ decays. The $B^0_s\to\mu^+\mu^-$ mode has been observed,
and $\Delta \Gamma_s$ provides access to a new  -- theoretically clean -- observable ${\cal A}_{\Delta\Gamma_s}^{\mu\mu}$, 
which should be fully exploited in the future. What are the implications of the $B^0_s\to\mu^+\mu^-$ measurement for the other 
$B_{s,d}\to \ell^+\ell^-$ decays? Assuming flavour-universal NP effects, $B_d\to\mu^+\mu^-$ is found to be moderately suppressed 
with respect to the SM and the NP effects strongly suppressed by $m_\mu/m_\tau\sim 0.06$ in $B_{s,d}\to\tau^+\tau^-$ decays. 
On the other hand, NP effects could by hugely amplified in this scenario by $m_\mu/m_e\sim 207$ in $B_{s,d}\to e^+e^-$, thereby 
lifting $\overline{\mathcal{B}}(B_s \rightarrow e^+e^-)$ up to the regime of $\overline{\mathcal{B}}(B_s \rightarrow \mu^+\mu^-)$, 
with the exciting possibility that it may be within reach at the LHC. New sources of CP violation may enter  $B_q\to\ell^+\ell^-$ 
decays and offer an interesting playground, both for theorists within specific extensions of the SM and for experimentalists to 
explore future measurements of the corresponding observables. Decays of the kind $B_q\to\ell^+\ell^-$ offer new degrees of 
freedom for NP searches at the upcoming LHC upgrade and beyond!

\section*{Acknowledgments}
I am very grateful to Gilberto Tetlalmatzi-Xolocotzi, Ruben Jaarsma and Daniela Gal\'arraga Espinosa for our recent collaborations
on topics discussed here. This research has been supported by the Netherlands Foundation for Fundamental Research of 
Matter (FOM) programme 156, ``Higgs as Probe and Portal'', and by the National Organisation for Scientific Research (NWO).

\section*{References}
\end{document}